\documentclass[twoside]{article}
\usepackage{fleqn,espcrc2}

\usepackage{epsfig}
\usepackage{latexsym,amsfonts}
\usepackage[figuresright]{rotating}

\newcommand{\AmS}{{\protect\the\textfont2
  A\kern-.1667em\lower.5ex\hbox{M}\kern-.125emS}}

\title{Understanding HTS cuprates based on the phase string theory of
doped antiferromagnet}
\author{Z.Y. Weng, D.N. Sheng, and C.S. Ting 
\address{Texas Center for Superconductivity,
University of Houston, Houston, TX 77204, U.S.A.}}
\begin{document}
\begin{abstract}
We present a self-consistent RVB theory which unifies the
metallic (superconducting) phase with the half-filling antiferromagnetic (AF)
phase. Two crucial factors in this theory include the RVB condensation which 
controls short-range AF spin correlations and the phase string effect 
introduced by hole hopping as a key doping effect. We discuss both the
uniform and non-uniform mean-field solutions and show the unique features of 
the characteristic spin energy scale, superconducting transition 
temperature, and the phase diagram, which are all consistent with the 
experimental measurements of high-$T_c$ cuprates.

\vspace{-1pc}
\end{abstract}

\maketitle

The low-energy physics of high-$T_c$ cuprates is essentially described by
a doped Mott insulator \cite{anderson}. The half-filling insulating phase can be 
well understood based on the Heisenberg model, where the bosonic 
resonating-valence-bond (RVB) state \cite {liang} gives a highly accurate 
description of both short-range and long-range spin correlations. And the 
mean-field bosonic RVB theory \cite{aa} provides a very useful analytic
framework over a wide range of temperature.

The real challenge comes from the doped case: How the doped holes will
influence the AF spin background and by doing so shape their own
dynamics. A simple generalization of the aforementioned AF mean-field theory 
\cite{aa} to finite doping has indicated that the motion of doped holes 
generally gets totally frustrated at the mean-field level such that an 
additional spiral spin twist must be induced in
order to gain a finite kinetic energy. But this is only an
{\it artifact} due to a mistreatment of the singular phase string
effect introduced by doped holes to be discussed below. We will show that the bosonic RVB ordering, which characterizes 
the short-range spin-spin correlations at half-filling, actually {\it favors}
the hopping of doped holes and the RVB order parameter will still control
the doped phase just like in the undoped case, a picture consistent with the 
original motivation for introducing the RVB concept by Anderson \cite{anderson}.

\indent {\bf PHASE STRING EFFECT.} The $t-J$ model in the Schwinger-boson, 
slave-fermion representation has the form $H_{t-J}=H_t + H_J$:
\begin{equation}  \label{et}
H_t=-t\sum_{\langle ij\rangle}\hat{H}_{ij}\hat{B}_{ji}+ H.c.,
\end{equation}
\begin{equation}  \label{ej}
H_J=-\frac J 2 \sum_{\langle ij\rangle}\left(\hat{\Delta}^s_{ij}\right)^{\dagger} \hat{\Delta}^s_{ij},
\end{equation}
where
$\hat{H}_{ij}=f^{\dagger}_if_j$,
$\hat{B}_{ji}=\sum_{\sigma}\sigma b^{\dagger}_{j\sigma}b_{i\sigma}$,
and $\hat{\Delta}^s_{ij}=\sum_{\sigma}b_{i\sigma}b_{j-\sigma}$.
Here $f_i$ is a fermionic ``holon'' operator and $b_{i\sigma}$
is known as the Schwinger-boson operator. At half-filling, only $H_J$ exists
and the mean-field state \cite{aa} is described by the bosonic RVB order
parameter $\Delta ^s=\langle \hat{\Delta}_{ij}^s\rangle $. Note that here in the 
slave-fermion decomposition, $c_{i\sigma}=f^{\dagger}_ib_{i
\sigma}(-\sigma)^i$, the Marshall
sign is explicitly built in through $(-\sigma)^i$ such that there is no sign
$\sigma$ appearing in $\hat{\Delta}^s_{ij}$. But $\hat{B}_{ji}$ in the
hopping
term acquires a sign $\sigma$, which would lead to $\langle\hat{B}
_{ji}\rangle=0$ in the RVB state, namely, the absence of the bare
hopping for doped holes at the mean-field level. 

However, if one follows the
hopping of a hole continuously without making the mean-field average at
first, then one finds \cite{string1} that a sequence of signs $(-1)\times
(+1)\times (+1)...=(-1)^{N_c^{\downarrow }}$ will be picked up 
by the hole along its path $c$, where $N_c^{\downarrow }$ counts the total
number of $\downarrow $ spins being {\it exchanged} with the given hole, 
originated from the simple sign $\sigma $ mentioned above.
To illustrate the importance of such a phase string effect, let us take the one
hole case as an example by expressing the total energy as follows
\begin{eqnarray}  \label{ge}
E_{\mathbf{k}}=E_0^J -\frac t N \sum_{ij}e^{-i\mathbf{k}\cdot (\mathbf{r}_i-
\mathbf{r}_j)} M_{ij}, \\ \nonumber
M_{ij}\equiv \sum_{\{c\}, N_c^{\downarrow}} P[\{c\}; N_c^{\downarrow}] (-1)^{N_c^{\downarrow}},
\end{eqnarray}
with using the Brillouin-Wigner formula, where $E_0^J$ denotes the ground-state
energy with the hole being \textit{fixed} at a given lattice site. The energy 
gain due to the hopping comes from the second term of (3).  In $M_{ij}$ the summation runs over all the possible paths $\{c\}$ of the hole connecting $i$ 
and $j$. The significant fact is that the weight functional
$P$ can be shown to be {\it positive-definite} at $
E_{\mathbf{k}}<E_0^J$. (The derivation is very similar to that of the
single-electron propagator in the one-hole case \cite{string1}  and the
details will be presented
elsewhere.) It implies that
the phase string factor $(-1)^{N_c^{\downarrow }}$ is the \textit{only}
source of phase frustration which cannot be ``repaired'' since $P\geq 0$, and
thus is expected to dominate the low-energy physics. The only additional
phase effect when there are many holes will come from the exchange of any
two holes due to the fermionic
nature of holons in the slave-fermion representation. At small doping,
the phase string effect obviously remains dominant.
  
Thus the conventional slave-fermion scheme as a very successful
framework at half-filling is no longer 
useful at finite doping if one does not know how to deal with the singular,
nonlocal phase string effect [which basically counts how many $\downarrow$ (or $\uparrow$ by symmetry) spins exchanged with the hole during its propagation) 
hidden in the formalism. On the other hand, 
since the phase string effect induced by doped holes is always present in the
ground state and can be ``counted'' in terms of the number of exchange between holons and 
spinons, it is possible to incorporate the singular effect explicitly into the
wavefunction like the fermion statistics. It turns out that one can realize
this through a unitary transformation \cite{string1} through which the $t-J$ 
Hamiltonian is ``regulated'' as the local singular sign $\sigma$ is ``gauged away''. The resulting exact
formalism is known as the phase string representation,
where the RVB order parameter becomes
\begin{equation}
\hat{\Delta}_{ij}^s=\sum_\sigma \left( e^{-i\sigma A_{ij}^h}\right) \bar{b}_{i\sigma}\bar{b}_{j-\sigma },  \label{delta}
\end{equation}
while in the hopping term (\ref{et}), $\hat{B}_{ji} \rightarrow
\hat{B}_{ji}=\sum_\sigma \left( e^{i\sigma A_{ji}^h}\right) \bar{b}_{j\sigma 
}^{\dagger}\bar{b}_{i\sigma }$
and $\hat{H}_{ij}=\left( e^{iA_{ij}^f}\right) h_i^{\dagger }h_j$.
Note that the fermionic operator $f_i$ now is replaced by a {\it bosonic}
holon operator $h_i$ after the transformation while $b_{i\sigma}\rightarrow 
\bar{b}_{i\sigma}$ which is still a boson operator. In this new ``bosonization'' formalism, even though the singular sign $\sigma$ in the 
original $\hat{B}_{ji}$ is ``gauged away'', the topological effect of the phase 
string (the total phase for a closed path) is precisely tracked by 
lattice gauge fields $A_{ij}^f$ and $A_{ij}^h$. Here
$A_{ij}^f\equiv A_{ij}^s-\phi_{ij}^0$ with $\sum_C A_{ij}^s= 1 /2 \sum_{l\in C}\left(\sum_{\sigma}\sigma
n_{l\sigma}^b\right)$ for a closed path $C$ and $\sum_{\Box}\phi_{ij}^0=\pi$
per plaquette. And $\sum_{C} A_{ij}^h=1/ 2 \sum_{l\in C} n_l^h$.
($n_{l\sigma}^b$ and $n_l^h$ are spinon and holon number operators,
respectively.) Here $A_{ij}^s$ and $A_{ij}^h$ describe
quantized flux tubes bound to spinons and holons, respectively.
And $\phi_{ij}^0$ describes a uniform flux with a strength $\pi$ per plaquette.

{\bf MEAN FIELD THEORY.} In the phase string formulation, the hopping term 
becomes finite as $\langle \hat{B}_{ji}\rangle\neq 0$ \cite{string3} when
$\Delta^s \neq 0$. Namely, the single RVB field $\Delta^s$ will still
be the only order parameter that controls both the undoped and doped phases. 
Note that $\Delta^s$ decides the short-range spin-spin correlations as shown 
in the relation $\langle {\bf S}_i\cdot {\bf S}_j\rangle=-1/2 |\Delta^s|^2$ for 
nearest-neighboring $i$ and $j$.

In the following let us focus on the so-called uniform-phase solution 
where holons and spinons are assumed to distribute homogeneously without the
phase separation. In the ground state, the bosonic holons will experience the
Bose condensation (BC) such that one may replace $A^h_{ij}$ by $\bar{A}^h_{ij}$ 
satisfying $\sum_{\Box}\bar{A}^h_{ij}=\delta \pi$ ($\delta$ is the doping concentration) and $A^f_{ij}\rightarrow -\phi^0_{ij}$ as $A^s_{ij}$ diminishes 
due to the pairing of spinons with a gap opening (see below). Then the 
mean-field state can be straightforwardly obtained.

Fig. 1 illustrates how the spin dynamics evolves into the metallic regime: the
spin dynamic susceptibility at ($\pi$,$\pi$), $\chi''_{\pi\pi}(\omega)$, is 
shown at $\delta=0$ and $\delta=1/7\approx 0.143$, respectively. Compared to  half-filling,
a resonance-like single peak emerges at a finite energy $E_g$ with the weight 
diminishes below it. The inset shows the doping-dependence of $E_g$ which peaks
at $\sim 0.4 J$ (independent of $t$) around $\delta\sim 0.2$, which is very 
close to the famous $41$$meV$ peak observed \cite{neutron1} in optimal $YBCO$ 
compound by neutron scattering (for $J\sim 100$$meV$). $E_g$ decreases 
monotonically with reducing $\delta$ in low-doping regime. 
\begin{figure}[b!]
\epsfxsize=5.8 cm
\centerline{\epsffile{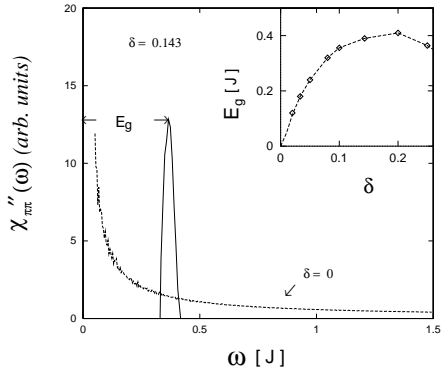}}
\caption{Dynamic spin susceptibility at the AF vector ($\pi$,$\pi$) as
the function of energy.}
\label{fig:1}
\end{figure}

The phase string effect is solely responsible for such a resonance-like peak
through the gauge field $\bar{A}^h_{ij}$. Another unique prediction of the 
phase string effect is shown in Fig. 2 for the local susceptibility $\chi_L''(\omega)=1/N\sum_{\bf q}\chi''({\bf q}, \omega)$ which shows multi-peak 
structure at higher energies like Landau levels as $\bar{A}^h_{ij}$ describes
uniform fictitious flux seen by spinons. Note that the high-energy peaks are 
contributed by those spin fluctuations away from the AF momentum ($\pi$,$\pi$)
in momentum space and are only significant under the integration over the whole 
Brillouin zone, which remains to be seen experimentally in the uniform phase 
like the optimal-$YBCO$ compound.     
\begin{figure}[t!]
\epsfxsize=5.8 cm
\centerline{\epsffile{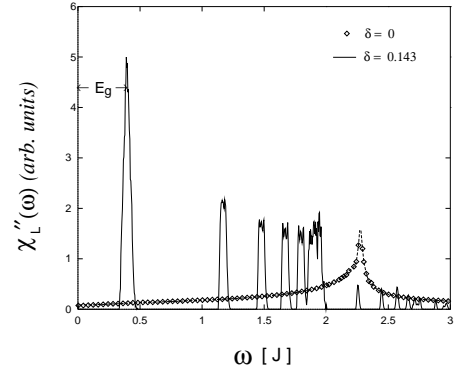}}
\caption{Local dynamic spin susceptibility as the function of energy} 
\label{fig:2}
\end{figure}

Another interesting property is that the ground state is intrinsically a 
superconducting state with d-wave symmetry. This can be 
verified by the paring order parameter
\cite{string3}
\begin{equation}  \label{sc}
{\Delta}^{SC}_{ij}= \Delta^s f_{ij} h^2_0 \left\langle e^{-\frac i
2 (\Phi^s_i+\Phi^s_j)}\right\rangle
\end{equation}
for nearest-neighboring sites $i$ and $j$. Here the numerical factor $f_{ij}$
changes sign from $j=i\pm \hat{x}$ to $j=i\pm \hat{y}$ determining the d-wave 
symmetry, $h_0=\langle h^{\dagger}_i\rangle$, and the phase $\Phi^s_i$ is defined 
as
\begin{equation}
\Phi_i^s=\sum_{l\neq i}\mbox{Im ln $(z_i-z_l)$}\sum_{\alpha}\alpha
n_{l\alpha}^b
\end{equation}
which describes vortices (anti-vortices) centered at up (down) spinons. The 
vortices-antivortices described by $\Phi^s_i$ are all paired up in
the ground state to ensure a finite $\Delta^{SC}_{ij}$. But at finite
temperature, free vortices appear in $\Phi^s_i$ as spinons are thermally excited 
will eventually kill the superconducting condensation. Using the effective
Hamiltonian for holon $H_h=-t_h\sum_{\langle ij\rangle 
}\hat{H}_{ij}+ h.c. $, $T_c$ can be estimated \cite{string3} as the temperature at which the excited spinon number equals to the total holon number. Fig. 3
show the correlation between $T_c$ and the spin characteristic energy $E_g$ which is consistent with the experimental relation for the cuprate superconductors \cite{neutron2,neutron3}. In particular, $T_c \sim 
100$$K$ at $\delta=0.143$ with $E_g\sim 40 $$meV$ without tuning parameters 
except for $J\sim 100$$meV$.
\begin{figure}[b!]
\epsfxsize=5.5 cm
\centerline{\epsffile{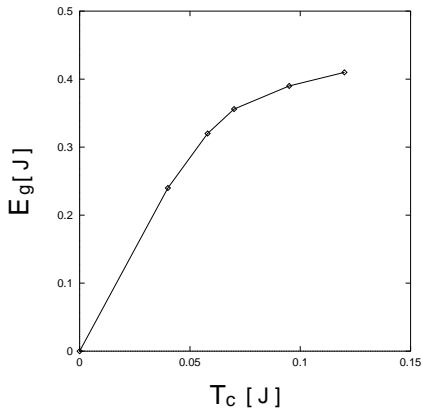}}
\caption{Relation between the spin energy $E_g$ (Fig.1) and $T_c$.} 
\label{fig:3}
\end{figure}

The phase diagram of the present mean-field theory is illustrated in Fig. 4.
Note that the whole theory is underpinned by the RVB order parameter $\Delta^s$
controlling the nearest-neighboring AF spin correlations which are gradually
reduced with the increase of $\delta$. Both AF long-range ordering in the 
insulating phase (holes must be localized) and superconducting condensation
in metallic phase occur inside such an RVB background, characterized by the BCs 
of spinons and holons, respectively. Note that there is an underdoped 
metallic regime denoted by SBC in Fig. 4 under the dashed curve 
which indicates the case of spinon BC in the metallic phase, corresponding to a non-uniform solution of the present mean-field
theory \cite{string3}. In the above uniform phase, the spin gap $E_g$ shown in 
Fig. 1 prevents the BC of spinons. But a 
non-uniform distribution of holons will lead to a spatial redistribution of the 
fictitious flux seen by spinons through $A^h_{ij}$ to generate a low-energy tail 
in spinon spectrum. At low doping, the Bose condensation of spinons can happen 
as the tail extends to zero, which physically means short-range AF orders 
exist in the hole-dilute regions. Whether such a solution is a micro or macro 
phase separation still needs further investigation. But generally the dynamic
spin susceptibility function will show a double-peak structure in energy space
near ($\pi$, $\pi$) with a pesudo-gap feature as there always will be some 
weight extends towards the zero energy limit \cite{string3}. This fact may have
already seen experimentally in the underdoped $YBCO$ compounds \cite{neutron2,neutron3}.    
\begin{figure}[t!]
\epsfxsize=5.2 cm
\centerline{\epsffile{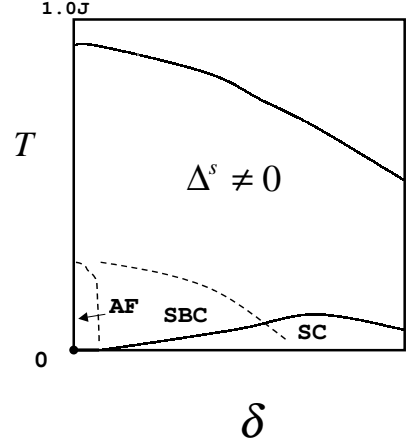}}
\caption{Phase diagram of the RVB state characterized by $\Delta^s$: AF and SC 
phases correspond to spinon and holon Bose condensations, respectively. 
And SBC denotes the spinon Bose condensation in the metallic regime.} 
\label{fig:4}
\end{figure}

\end{document}